\documentstyle[12pt,aaspp4]{article}
\def\lsim{\lower.5ex\hbox{$\; \buildrel < \over \sim \;$}}
\def\gsim{\lower.5ex\hbox{$\; \buildrel > \over \sim \;$}}
\lefthead{}
\righthead{Stability of accretion disk around rotating black holes}

\def\lsim{\lower.5ex\hbox{$\; \buildrel < \over \sim \;$}}
\def\gsim{\lower.5ex\hbox{$\; \buildrel > \over \sim \;$}}

\received{2002 September 8}
\begin{document}

\title{Stability of accretion disk around rotating black holes: a pseudo-general-relativistic
fluid dynamical study}
 
\author{Banibrata Mukhopadhyay}

\affil{\small Inter-University Centre for Astronomy and Astrophysics,
Post Bag 4, Ganeshkhind, Pune-411007, India\\
}

\begin{abstract}

We discuss the solution of accretion disk when the black hole is chosen
to be rotating. We study, how the fluid properties get affected for different
rotation parameters of the black hole. We know that no cosmic object is static in 
Universe. Here the effect of the rotation of the black hole to the space-time
is considered following an earlier work of the author, where the pseudo-Newtonian potential was
prescribed for the Kerr geometry. We show that, with the inclusion of rotation of
the black hole, the valid disk parameter region dramatically changes and disk becomes
unstable. Also we discuss about the possibility of shock in accretion disk around
rotating black holes. When the black hole is chosen to be rotating, the sonic locations of the
accretion disk get shifted or disappear, making the disk unstable.
To bring it in the stable situation, the angular momentum 
of the accreting matter has to be reduced/enhanced (for co/counter-rotating disk)
by means of some physical process. 

\end{abstract}


\section{Introduction}

For over last three decades, fluid dynamical studies of the accretion disk around black holes have
been extensively studied. Shakura \& Sunyaev (1973) initiated this discussion considering 
very simplistic but effective model of accretion disk. They chose the Newtonian gravitational potential.
Novikov \& Thorne (1973) and Page \& Thorne (1974) studied a few aspects of the accretion disk in a
full relativistic treatment. Paczy\'nski \& Wiita (1980) proposed one pseudo-potential which 
could approximately describe relativistic properties of the accretion disk around non-rotating 
black holes. Using that pseudo-potential, Abramowicz \& Zurek (1981) studied some transonic 
properties of accretion flows. Later on, with time, number of studies of accretion disk
have been carried out with modified disk structures (e.g. Abramowicz et al. 1988; Abramowicz \& Kato 1989; 
Chakrabarti 1990, 1996a; Narayan \& Yi 1994; Narayan et al. 1997, 1998;  Kato et al. 1998;
Mukhopadhyay 2002a) for non-rotating black holes. Also the studies of accretion disk
in Kerr geometry have been explored by several authors (Sponholz \& Molteni 1994;
Chakrabarti 1996b, 1996c; Abramowicz et al. 1996; Peitz \& Appl 1997; Gammie \& Popham 1998; 
Popham \& Gammie 1998; Miwa et al. 1998; Lu \& Yuan 1998; Manmoto 2000). Those studies have 
been made either in a full general relativistic or in a pseudo-Newtonian approach. 

The study on stability of accretion disk is one of the important criteria in this context. 
The possibility of steady and stable disk formation by incoming matter towards a black
hole is allowed only for certain sets of initial parameters. Abramowicz \& Zurek (1981)
studied the effect of angular momentum on the accretion and corresponding stability
of the transonic nature of the infalling matter onto the black hole. After that,
Chakrabarti (1989, 1990) discussed about the formation
of sonic points and shock waves in accretion disks. He showed that the formation of
stable sonic points and shock in the disk are possible for a particular range of 
physical parameters.
Recently, Mukhopadhyay \& Chakrabarti (2001) have analysed the stability of accretion 
disk in presence
of nucleosynthesis. As the temperature of the disk is very high (at least of the order 
of $10^9$K), nuclear burning can take place and generate or/and absorb energy in a 
large scale in the disk. Because of this generation/absorption of nuclear energy, 
fundamental properties of the disk, like sonic locations, temperature,
Mach number, shock location (if any), etc., may be affected. Thus the stability gets 
influenced by the nuclear burning. If the rate of generation/absorption of nuclear 
energy is large, the disk structure may become unstable. In their discussion, 
Mukhopadhyay \& Chakrabarti (2001) have also shown
that the sonic locations disappear for a particular shell of accretion disk where the
nuclear energy is significant compared to the mechanical energy of the accreting matter. 

In all the above mentioned works, when the stability was analysed for the accretion disk, the central black
hole was chosen to be non-rotating. As it is well known that no object is static in the Universe, 
before making any serious conclusions about the inner properties of the accretion disk, the consideration
of rotation of a black hole is essential. Though a few works have been done
for Kerr black holes, no such comparative study has been explored so far with different Kerr parameters. 
Thus automatically the following
questions arise, does the rotation of black hole play an important role in the accretion properties of a
disk? Does the consideration of rotation of black hole affect the disk structure for a particular 
parameter regime which is known 
to be stable for the non-rotating case? Here, we would like to address all these questions one by one. 

Recently, we have proposed a new pseudo-Newtonian potential (Mukhopadhyay 2002b, hereinafter Paper-I) 
to describe the accretion disk around Kerr black holes. Following Paper-I, we will describe our accretion 
disk in a pseudo-Newtonian manner. The potential, that we are using, can describe approximately 
all the essential general relativistic properties of the accretion disk. 
It can reproduce the locations of marginally stable and
bound orbit exactly or almost exactly for different Kerr parameters (see Paper-I), 
which are the pure general 
relativistic properties. Therefore, we call this study as a {\it pseudo-general-relativistic approach}. 
As our main interest is to study the inner region of the accretion disk, we will concentrate upon 
the sub-Keplerian flow, where the effect of rotation of black hole is 
significant. In Paper-I, we already raised several questions, which are important to address 
in physics of accretion disk. Here we will show, how useful is the pseudo-potential prescribed
in Paper-I to study the global behaviour of the accretion disk. 
In full general relativity, the basic equations of accretion disk are very complicated and
tedious to handle. In Early, there were no suitably good pseudo-potential for a rotating black hole.
It seems that Paper-I has come up with the answer and one of our aim in this paper is to show its applicability.
As the location of horizon, marginally bound and marginally stable orbits change for different 
Kerr parameters, any inner property of the disk is very much related to the rotation of black hole.
We will follow Mukhopadhyay \& Chakrabarti (2001) to analyze the stability of the disk. We will study, 
whether or not the sonic location(s), shock formation(s) (if any), etc., gets affected by applying 
rotation to the black hole.
In an accretion disk, if the sonic point looses entropy or disappears by the rotational effect 
of black hole, we will understand that the sonic point as well as the 
disk is unstable because stable sonic point is necessary for the accretion onto black hole.
In the next section, we will present the basic equations needed to describe the accretion disk. 
In \S 3, we will discuss about the parameter space of accretion disk and how does the rotation affect it.
Subsequently, in \S 4, we will describe the fluid dynamical results. Finally, in
\S 5, we will sum up all the results and make the overall conclusions.

\section{Basic Equations }

Here, throughout in our calculations, we express the radial coordinate
in unit of $GM/c^2$, where $M$ is mass of the black hole, $G$ is the gravitational constant and
$c$ is the speed of light. We also express the velocity in unit of speed of
light and the angular momentum in unit of $GM/c$. The equations to be solved are given below as:
   \begin{equation}
   \frac{d}{dx}(x\Sigma v)=0,
   \label{ec}
   \end{equation}
   \begin{equation}
   v\frac{dv}{dx}+\frac{1}{\rho}\frac{dP}{dx}-\frac{\lambda^2}{x^3}+F(x)=0 
   \label{rmom}
   \end{equation}
   where, $\Sigma$ is the vertically integrated density and $F(x)$ is the gravitational pseudo-Newtonian
   force given in Paper-I as
   \begin{equation} 
   F(x)=\frac{(x^2-2a\sqrt{x}+a^2)^2}{x^3(\sqrt{x}(x-2)+a)^2},
   \label{pf}
    \end{equation}
where $a$ indicates the specific angular momentum of the black hole (Kerr parameter). 
We do not consider any kind of energy dissipation in the accretion disk, as we want to check, how the black
hole rotation solely can affect the disk properties. Thus the angular momentum of accreting fluid ($\lambda$)
remains constant throughout a particular flow. Following Matsumoto et al. (1984), we can calculate the 
vertically integrated density as
\begin{equation}
  \Sigma=I_n\rho_e h(x), 
  \label{den}
  \end{equation}
  where, $\rho_e$ is the density at equatorial plane, $h(x)$ is the half-thickness of the disk and
  $I_n=\frac{(2^n n!)^2}{(2n+1)!}$, where $n$ is the polytropic index. From the vertical equilibrium 
  assumption, the half-thickness can be written as
  \begin{equation}
   h(x)=c_s x^{1/2}F^{-1/2},
  \label{ht}
  \end{equation}
  where $c_s$ is the speed of sound. We also consider the equation of state to be $\gamma P/\rho=c_s^2$, where
  $\gamma$ is the gas constant.

Now, combining (\ref{ec}) and (\ref{rmom}) we get
{\large
\begin{equation}
\frac{dv}{dx}=\frac{\frac{\lambda^2}{x^3}-F(x)+\frac{c_s^2}{\gamma+1}\left(\frac{3}{x}-\frac{1}{F}\frac{dF}{dx}\right)}
{v-\frac{2c_s^2}{(\gamma+1)v}}.
\label{dvdx}
\end{equation}
}
Far away from the black hole, $v<c_s$ and close to it, $v>c_s$, thus there is an intermediate
location where the denominator of (\ref{dvdx}) must vanish. Therefore, to have a smooth solution at that
location, the numerator has to be zero. This location is called the sonic point or critical point ($x_c$). The existence of 
this sonic location plays an important role in the accretion phenomena. For an accretion disk around black 
hole, sonic point must exist. From the global analysis of sonic point, one can 
understand the stability of physical parameter region which we will discuss in the next section. 

As it is described above that, at $x=x_c$, $\frac{dv}{dx}=\frac{0}{0}$, using l'Hospital's rule
and after some algebra, we can get the velocity gradient of accreting matter at the sonic
location as
\begin{equation}
\frac{dv}{dx}|_c=-\frac{{\cal B}+\sqrt{{\cal B}^2-4{\cal A}{\cal C}}}{2{\cal A}},
\label{dvdxc}
\end{equation}
where, 
\begin{eqnarray}
\nonumber
{\cal A}&=&1+\frac{2c_{sc}^2}{(\gamma+1)v_c^2}+\frac{4c_{sc}^2(\gamma-1)}{v_c^2(\gamma+1)^2},\\
\nonumber
{\cal B}&=&\frac{4c_{sc}^2(\gamma-1)}{v_c(\gamma+1)^2}\left(\frac{3}{x_c}-\frac{1}{F_c}\frac{dF}{dx}|_c\right),\\
{\cal C}&=&\frac{c_{sc}^2}{\gamma+1}\left[\left(\frac{1}{F_c}\frac{dF}{dx}|_c\right)^2-\frac{1}{F_c}
\frac{d^2F}{dx^2}|_c-\frac{3}{x_c^2}\right]-\frac{3\lambda^2}{x_c^4}-\frac{dF}{dx}|_c\\
\nonumber
&-&\frac{c_{sc}^2(\gamma-1)}{(\gamma+1)^2}\left[\frac{3}{x_c}-\frac{1}{F_c}\frac{dF}{dx}|_c\right]^2.
\label{cof}
\end{eqnarray}
Thus, we have to integrate (\ref{dvdx}) and (\ref{dvdxc}) with an appropriate boundary condition to 
get the fluid properties in accretion disk. From (\ref{dvdx}), we can easily find out the Mach number
at the sonic point as
\begin{equation}
{\it M}_c=\frac{v_c}{c_{sc}}=\sqrt{\frac{2}{\gamma+1}}
\label{mac}
\end{equation}
and the corresponding sound speed as
\begin{equation}
c_{sc}=\sqrt{(\gamma+1)\left(\frac{\lambda^2}{x_c^3}-F_c\right)\left(\frac{1}{F_c}\frac{dF}{dx}|_c
-\frac{3}{x_c}\right)^{-1}}.
\label{csc}
\end{equation}
Now integrating (\ref{ec}) and (\ref{rmom}), we can write down the energy and entropy of the flow at sonic point as 
\begin{equation}
E_c=\frac{2\gamma}{(\gamma-1)}\left(\frac{\frac{\lambda^2}{x_c^3}-F_c}{\frac{1}{F_c}\frac{dF_c}{dx}|_c
-\frac{3}{x_c}}\right)+V_c+\frac{\lambda^2}{2x_c^2},
\label{Ec}
\end{equation}
and
\begin{equation}
\dot{\cal M}_c=(\gamma K)^n\dot{M}=x_c^{3/2} F_c^{-1/2}(\gamma+1)^{q/2}\left(\frac{\frac{\lambda^2}{x_c^3}-F_c}{\frac{1}{F_c}
\frac{dF}{dx}|_c-\frac{3}{x_c}}\right)^{\frac{\gamma}{\gamma-1}},
\label{muc}
\end{equation}
where $K$ is the gas constant that is basically the entropy of the system, $V_c=(\int F dx)|_c$ 
and $q=\frac{(\gamma+1)}{2(\gamma-1)}$. Actually $\dot{\cal M}$ itself is not entropy but
carries the information of it. For a non-dissipative system, 
$\dot{\cal M}$ is conserved for a particular $\dot{M}$, unless the shock forms there.
We have to supply the sonic energy $E_c$ as the boundary condition for a particular flow. 
Then from (\ref{Ec}), we can find out the sonic location $x_c$. Therefore, knowing $x_c$, one can
easily find out the fluid velocity and sound speed at the sonic point from (\ref{mac}) and (\ref{csc}) 
for a particular accretion flow. These have to be supplied as further boundary conditions of the flow.

Another important issue is the formation of shock in the discussion of accretion disk.
Here, whenever we mention the shock, will mean the Rankine-Hugoniot shock (Landau \& Lifshitz 1959).
If we generalize the conditions to form a shock in accretion disk given by Chakrabarti (1989), to the
case of a rotating black hole, we get
\begin{equation}
\frac{1}{2}M_+^2 c_{s+}^2+nc_{s+}^2=\frac{1}{2}M_-^2 c_{s-}^2+nc_{s-}^2,
\label{scke}
\end{equation}
\begin{equation}
\frac{c_{s+}^\nu}{\dot{\cal M}_+}\left(\frac{2\gamma}{3\gamma-1}+\gamma M_+^2\right)
=\frac{c_{s-}^\nu}{\dot{\cal M}_-}\left(\frac{2\gamma}{3\gamma-1}+\gamma M_-^2\right),
\label{sckmom}
\end{equation}
\begin{equation}
\dot{\cal M}_+>\dot{\cal M}_-,
\label{sckent}
\end{equation}
where
\begin{equation}
\dot{\cal M}=Mc_s^{2(n+1)}\frac{x_s^{3/2}}{\sqrt{F(x_s)}}.
\label{sckent1}
\end{equation}
Here, subscripts '$-$' and '$+$' indicate the quantities just before and after the shock 
respectively and $x_s$ indicates the shock location. $M$ denotes 
the Mach number of the matter and $\nu=\frac{3\gamma-1}{\gamma-1}$.
From (\ref{scke})-(\ref{sckent1}), it is very clear that except the entropy expression, all remain unchanged
with respect to the case of a non-rotating black hole. Also from (\ref{sckmom}) and (\ref{sckent1}), we get
the shock invariant quantity as
\begin{equation}
C=\frac{\left(\frac{2}{M_+}+(3\gamma-1)M_+\right)^2}{M_+^2(\gamma-1)+2}=
\frac{\left(\frac{2}{M_-}+(3\gamma-1)M_-\right)^2}{M_-^2(\gamma-1)+2}
\label{sckinv}
\end{equation}
which remains unchanged with respect to a non-rotating case. If all the conditions 
(\ref{scke})-(\ref{sckent}) and (\ref{sckinv}) are simultaneously satisfied by the matter, shock will only
form in an accretion disk.

\section{Analysis of the parameter space}

One of our aim is to check, how does the rotation affect the disk parameter region known
for Schwarzschild black hole. Therefore, we have to analyse globally the accretion disk properties
around rotating black holes. We also have to check, how does the rotation of black hole affect 
the sonic location, structure as well as the stability of disk.
In Fig. 1a, we show the variation of disk entropy as a function of sonic location.
The intersections of all the curves by the horizontal line (which is a constant entropy line)
indicate the sonic points of the accretion disk for that particular entropy and rotation
of the black hole. It is clearly seen that, at a particular entropy, if the rotation of black hole
increases, marginally bound ($x_b$) and stable ($x_s$) orbits as well as sonic points shift to a more 
inner region and the possibility to have all four sonic points in the disk outside the horizon
increases. As an example, for $a=0.998$, the inner edge of accretion disk 
enlarges in such a manner that the fourth sonic point in the disk appears outside the horizon. 
On the other hand, for retrograde orbits (counter rotating cases), $x_b$ and $x_s$ move to greater
radii and all the sonic points come close to each other for a particular Kerr parameter.
Similar features are reflected from Fig. 1b, where the sonic energy is plotted as a function of sonic
location. Here also, the intersections of the horizontal line (which indicates a constant energy line) 
by all the curves indicate the sonic points of the accretion disk for that particular energy and rotation
of the black hole. For both the Fig. 1a and 1b, sonic points with negative slope of the curve indicate the
locations of 'saddle-type' sonic point and positive slopes indicate the 'centre-type' sonic point.
Thus the rotation of black hole plays an important role to the formation and location of sonic points
which are related to the structure of accretion disk.

\clearpage

\begin{figure}
\epsscale{.80}
\plotone{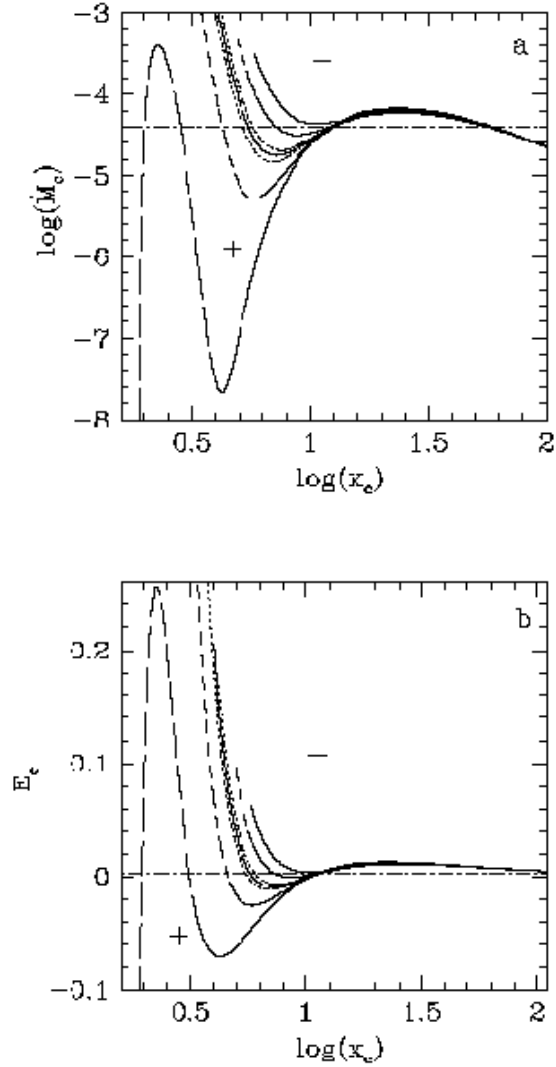}
\caption{
\label{fig1} Variation of (a) entropy and (b) energy as a function of sonic location for various values of 
Kerr parameters $a$. Central solid line indicates non-rotating case ($a=0$), while the results in regions of either side of it 
indicated by '+'  and '-' are for prograde and retrograde orbits respectively. Different 
curves from central solid line to downwards are for $a=0.1,0.5,0.998$ and to upwards for $a=-0.1,-0.5,-0.998$. 
The horizontal line indicates the curve of (a) constant entropy of $5\times 10^{-5}$ and (b) 
constant energy of $0.0065$. $\lambda=3.3$, $\gamma=4/3$ for all the curves. 
}
\end{figure}

\begin{figure}
\epsscale{0.8}
\plotone{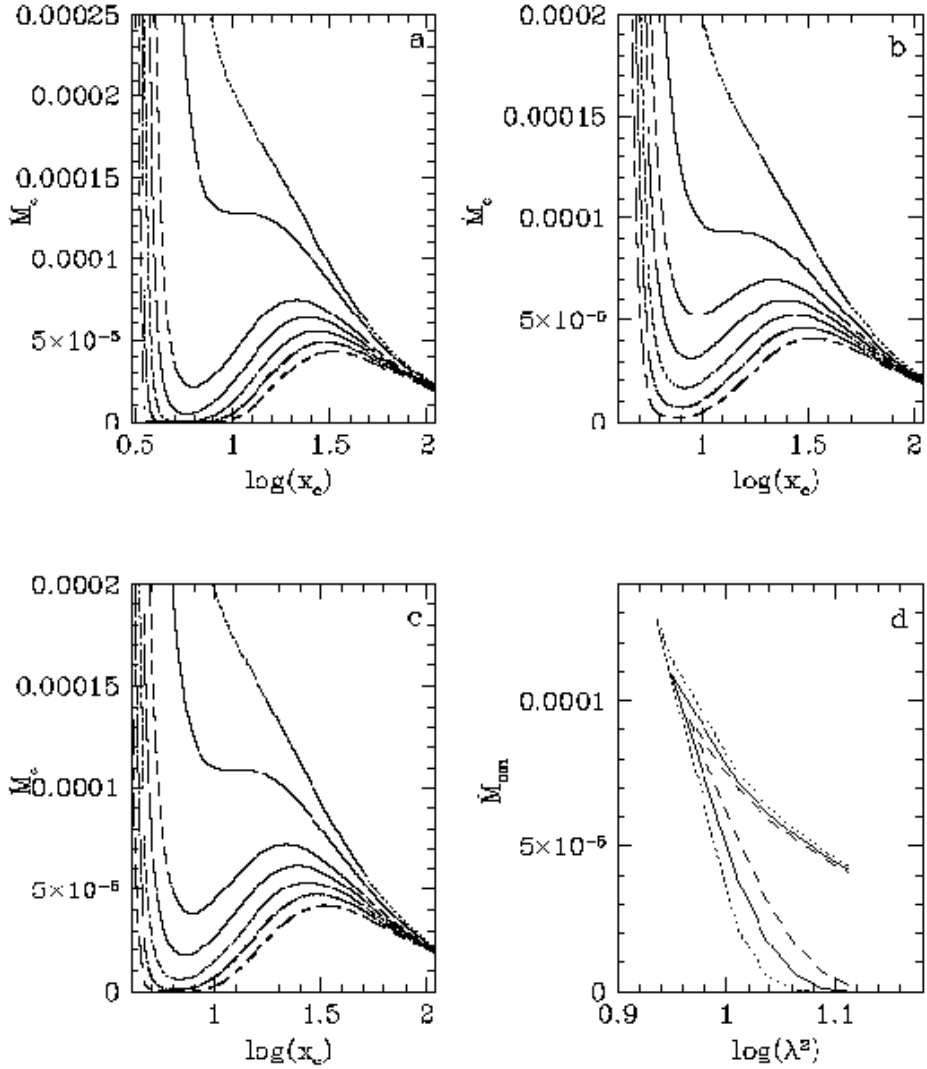}
\caption{
\label{fig2} (a-c) Variation of entropy as a function of sonic location for a set of disk angular momentum ($\lambda$), 
when (a) $a=0.5$, (b) $a=-0.5$ and (c) $a=0$. Solid curve indicates the flow with critical angular momentum ($\lambda_c$) 
for which the flow has just a single critical point which is (a) $\lambda_c=2.94$, (b) $\lambda_c=3.04$
and (c) $\lambda_c=2.982$. From the top to bottom curve, $\lambda=2.8,\lambda_c,3.2,3.3,3.4,3.5,3.6$. 
Any flow of $\lambda>\lambda_c$ has three and $\lambda<\lambda_c$ has one critical point.
(d) Variation of extremum values of entropy for curves (a), (b) and (c) as a function of $\lambda^2$ for
$a=0$ (solid curve), $a=0.5$ (dotted curve) and $a=-0.5$ (dashed curve). $\gamma=4/3$ for all the curves.}
\end{figure}

\clearpage

In Fig. 2, with the consideration of rotation of black hole, we show how the disk region having 
three sonic points is affected. Figure 2a,b,c show the variation of sonic entropy as a function of sonic
point, when the angular momentum of accreting matter is chosen as parameter for three 
different rotations of the black hole as $a=0.5,-0.5,0$ respectively. 
If we join the extrema of the curves, we get the bounded parameter region of the
disk where the three sonic points exist. As the angular momentum of accreting matter reduces,
the possibility of forming three sonic points reduces. This is very well-understood physically; as the
angular momentum of accreting matter reduces, disk tends to a {\it Bondi-Flow like} structure, that has 
single saddle-type sonic point.
The solid curves of all three figures indicate the locus of sonic entropy with a single extremum point. 
Only the flows of angular momentum greater than that of the solid curve have three sonic 
points, otherwise for a particular flow, we have only one sonic point.
In Fig. 2d, we show the projection of all the curves in Fig. 2a,b,c to the
$\dot{\cal M}_{cm}-\lambda^2$ plane, where $\dot{\cal M}_{cm}$ indicates the extremum 
entropy at the sonic points. Figure 2d indicates the valid physical parameter region of the accretion
disk that has three sonic points in the flow. It should be noted that, for a particular Kerr parameter,
the accreting flow can have three sonic locations for different pairs of
($\lambda, \dot{\cal M}_{cm}$). 
However, for a particular $\lambda$, if $\dot{\cal M}_{cm}$ decreases the stability of the 
sonic points as well as that of the disk diminishes too. Thus, in a section of the diagram 
for particular $a$, all points
are not equally probable set of physical parameters in the formation of three sonic points. 
It is clear that, as the co-rotation of black hole increases, the parameter region enlarges, where three 
sonic points may exist. Therefore, the disk with three sonic points becomes more probable at high co-rotation. 
On the other hand, this region reduces for the counter-rotating flow. Similar features
are noticed from Fig. 3, where the variation of sonic energies are depicted. In a similar manner,
Fig. 3d indicates that the parameter space with three sonic points enlarges if the co-rotation of 
black hole increases, where $E_{cm}$ indicates the extremum energy 
of the different curves in Fig. 3a,b,c. The physical reasons are the following. Higher co-rotation results in a 
 shift of $x_b$ and $x_s$ to a more inner edge of the accretion disk and an 
enlargement of the stable disk region (as we know that the stable circular orbit is possible in the disk 
upto the radius $x_s$). On the other hand, 
higher counter-rotation results in an outward shift of $x_b$ and $x_s$ and the size of the stable disk 
region reduces. 

\clearpage
\begin{figure}
\epsscale{0.9}
\plotone{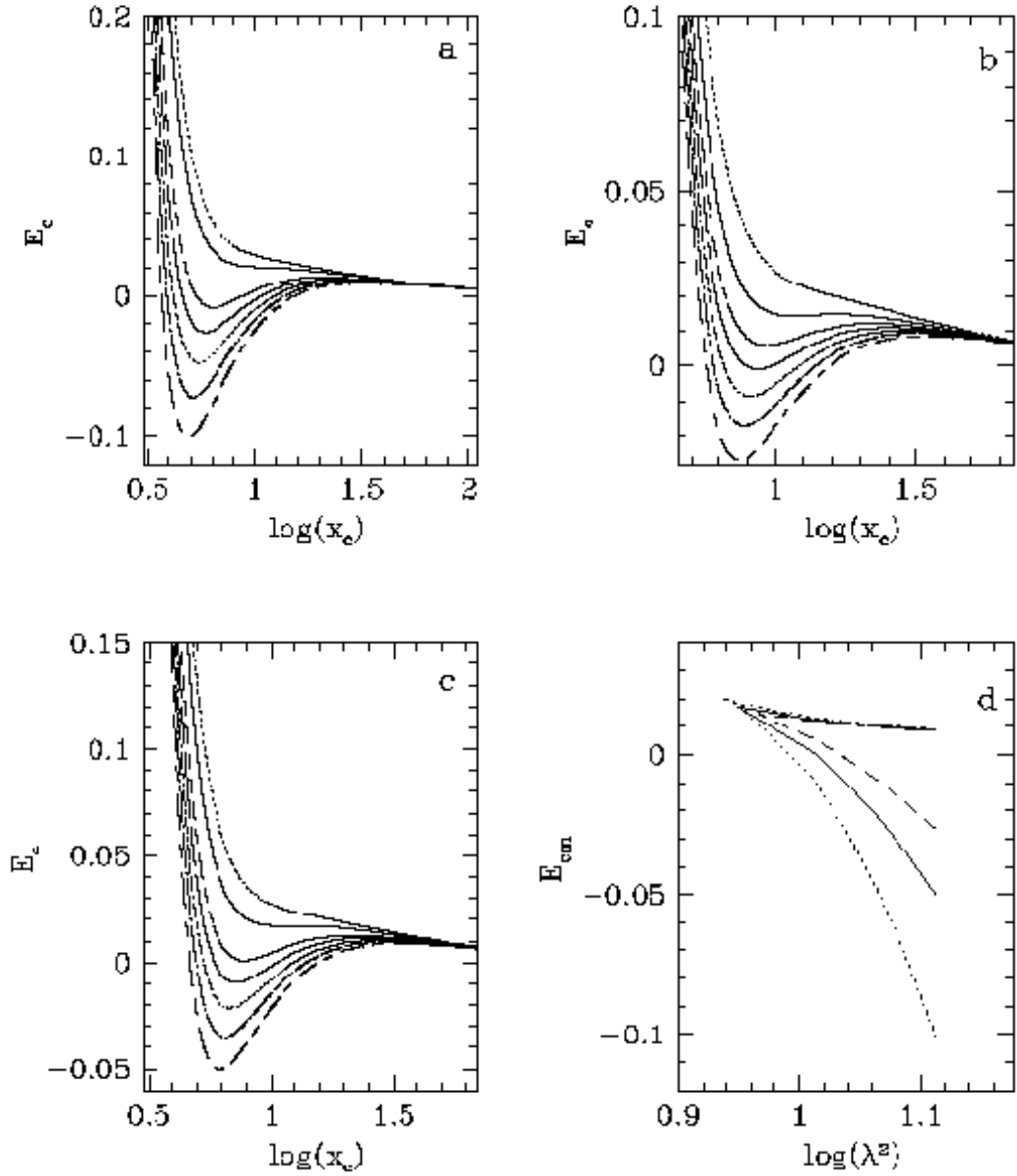}
\caption{
\label{fig3} Same as \ref{fig2} except the energy is plotted in place of entropy.}
\end{figure}
\clearpage

Figure 4 indicates the variation of energy with entropy at sonic locations for different
Kerr parameters. 'I' indicates the branch of inner 'X-type' sonic point that is different for
different Kerr parameters, while 'O' indicates the outer 'X-type' sonic point branch. Along
the branch O, all the curves (solid, dotted and dashed) lie on top of each other. Therefore it
is clear that the rotation of
black hole does not affect the outer 'X-type' sonic point.
Figures 4a,b show that with the increase of Kerr parameter (co-rotation), inner sonic
points of the disk shift towards the lower entropy region and the disk becomes unstable. But the outer 
sonic point branch remains unchanged for all values of rotation. We know that,
if there is a possibility of matter in the outer sonic point branch of lower entropy to jump to 
the inner sonic point branch of higher entropy, the shock can form in accretion disk. As, with 
the increase of Kerr parameter, the inner sonic point branch shifts to the lower entropy
region, the stability of both the shock and the disk is at stake. For example, in 
Fig. 4a, if we join the points A, B and A, D, we will find two sections of the curve namely, ABC 
(for solid curve) and ADE (for dashed curve) which are the parameter regions where the shock may 
form in accretion disk for $a=0$ and $a=0.5$ respectively. 
If we draw the lines of constant energy (which are horizontal in Fig. 4a), connecting
the outer sonic point branch (AC for $a=0$ and AE for $a=0.5$) to the inner one (BC for $a=0$ and DE 
for $a=0.5$), section ABC will contain more number of lines with respect to that
in section ADE (as the section ABC is bigger than ADE). 
On the other hand, a constant energy line indicates the possibility of
shock (as it shows the possibility of increment of entropy in the flow, keeping the energy constant).
Clearly, for a co-rotating black hole, the possibility of formation of shock reduces as 
well as the shock itself becomes unstable. The physical reason is that, with the increase of Kerr
parameter, the angular momentum of the system increases, which helps the disk to maintain
a high azimuthal speed of matter upto a very close proximity to the black hole, as a result the radial matter speed can 
overcome the corresponding sound
speed at an inner radius only. On the other hand, the inner edge of the accretion disk is comparatively
less stable as the entropy decreases at lower radii. As the inner sonic points form at 
an inner radii, 
with the increase of $a$, the disk tends to an unstable situation 
that decreases the value of entropy.
Figure 4b shows that the possibility of shock
disappears completely when $a=0.998$ as there is no transition possible from the outer to inner sonic branch
that can increase entropy of the system. Here, the inner edge is extremely unstable in such a manner that 
its entropy is even lower than that of outer sonic point branch.
As the inner edge of the disk spreads out for $a=0.998$, there are two branches of the centre-type sonic. 

\clearpage
\begin{figure}
\epsscale{0.85}
\plotone{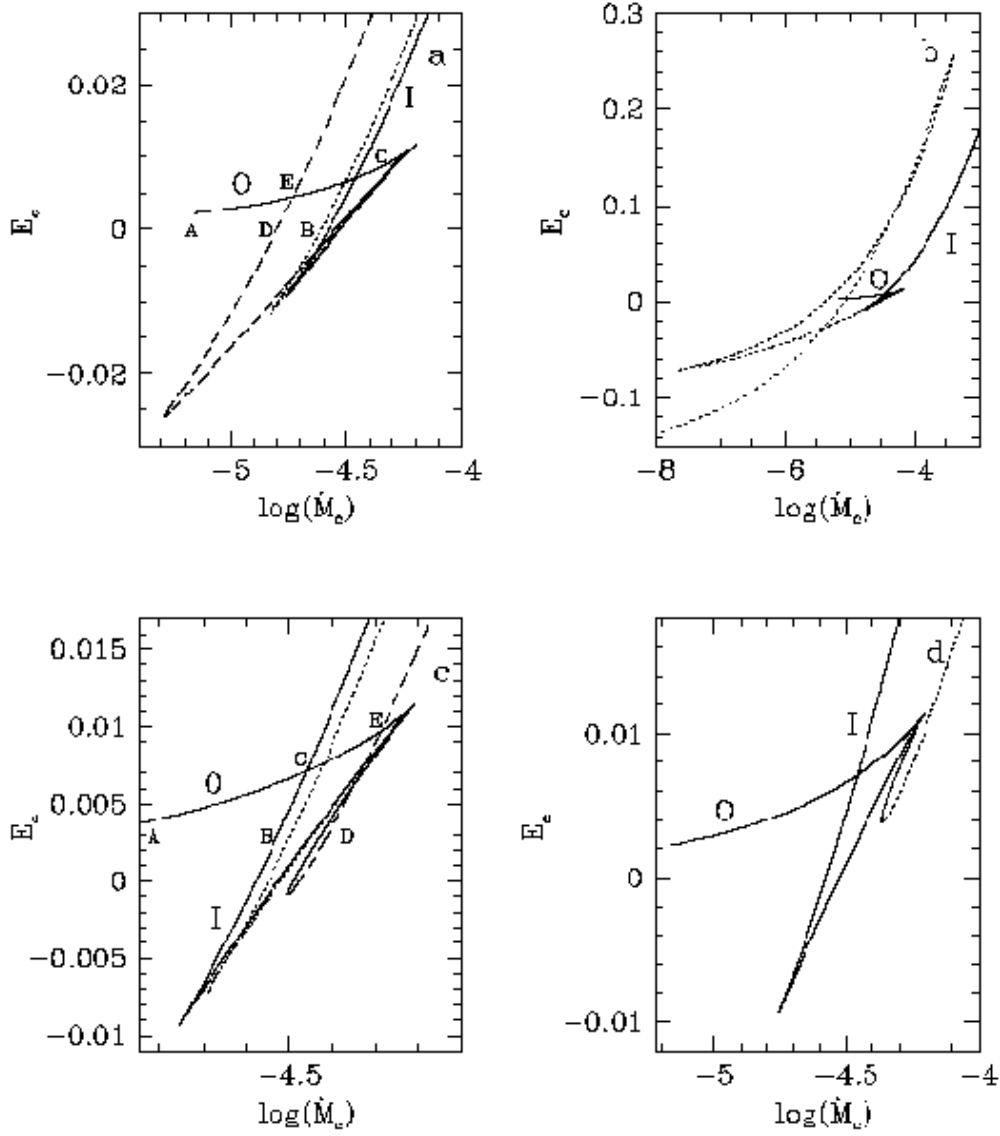}
\caption{
\label{fig4} Variation of energy with entropy at sonic points when disk angular momentum $\lambda=3.3$ for 
(a) $a=0$ (solid curve), $a=0.1$ (dotted curve) and $a=0.5$ (dashed curve), (b) $a=0$ (solid curve) and $a=0.998$ 
(dotted curve), (c) $a=0$ (solid curve), $a=-0.1$ (dotted curve) and $a=-0.5$ (dashed curve), 
(d) $a=0$ (solid curve) and $a=-0.998$ (dotted curve). O and I indicate the branch of outer and inner sonic 
points respectively for $a=0$. $\gamma=4/3$, for all the curves.}
\end{figure}
\clearpage

In Fig. 4c, the situation is opposite. As the counter-rotation of black hole increases in magnitude, 
the inner sonic point branch shifts towards a higher entropy region. Thus, the disk as well as the 
shock become more stable. Still, the outer sonic point branch is not affected due to the rotation
of black hole and the branches for different Kerr parameters are superimposed. 
In Fig. 4c, the sections ABC (for solid curve) and ADE (for dashed curve)
are for $a=0$ and $a=-0.5$ respectively, show the
regions where the shock may form in the disk. The physical reason to have a more stable inner sonic
point branch for the counter-rotating case is as follows. Higher the counter rotation 
implies that the net angular momentum of the system reduces,
as a result, the matter falls faster at larger radii and it
becomes supersonic at a comparatively outer edge of the disk that is more
stable. In a similar manner, as explained for Fig. 4a, the greater size of the section ADE than 
that of ABC implies the greater possibility of shock. However, for a very high 
counter-rotation of the black hole (see Fig. 4d), the entire system shifts to the outer 
side that is more stable, and the inner sonic point branch itself tends to merge to the branch of 
outer sonic point. Since the net angular momentum of the system becomes very small for
higher counter-rotation, the corresponding
centrifugal pressure onto the matter becomes insignificant, as a result the possibility of shock 
formation diminishes again. Therefore we can conclude that, if the angular momentum of black
hole increases or decreases significantly, the shock wave in accretion disk becomes unstable and 
the disk itself looses stability.

\section{Fluid Properties of Accretion Disk }

We will discuss now the effect of rotation of black hole on the disk fluid. 
In the previous section, we have already confirmed that the rotation significantly affects the 
global parameter space of accretion disk. Our intention is, to incorporate the rotation
of black hole and to see the change of fluid properties known for the non-rotating case.

\clearpage
\begin{figure}
\epsscale{0.8}
\plotone{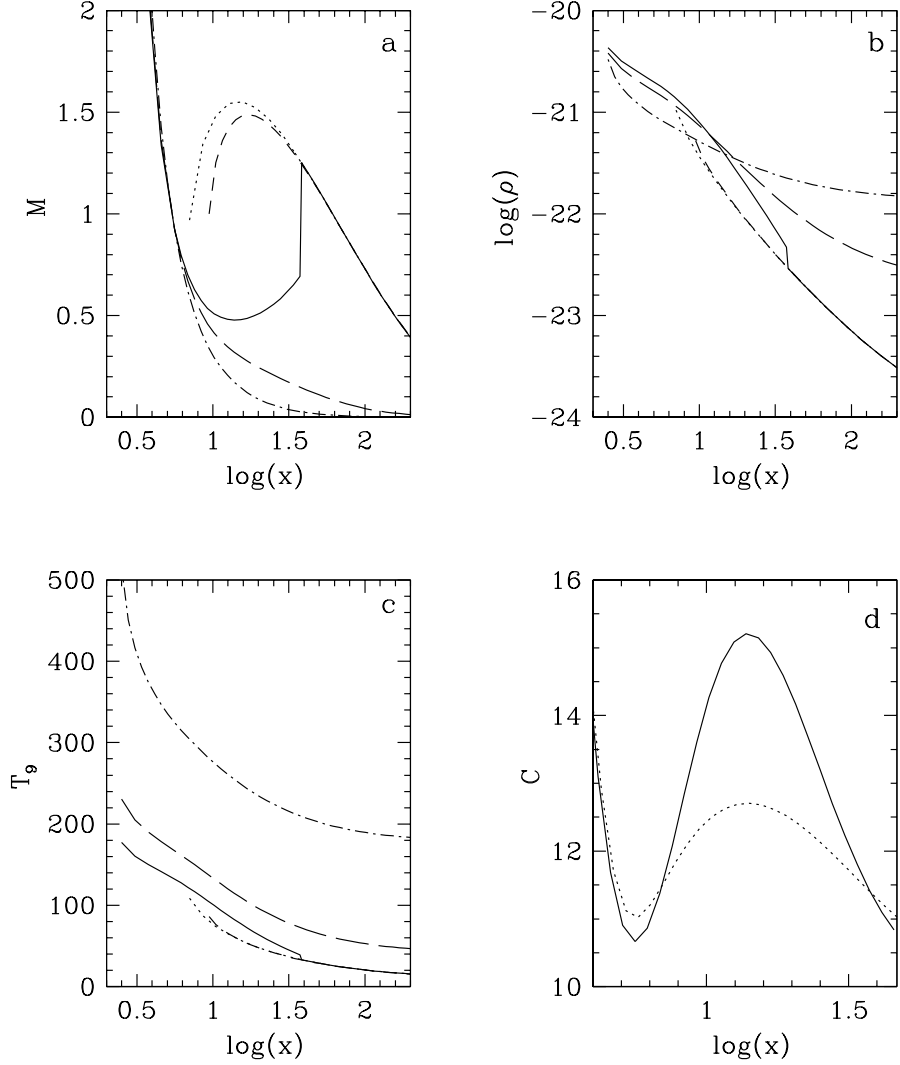}
\caption{
\label{fig5} Variation of (a) Mach no., (b) density and (c) temperature in unit of $10^9$ as a function of radial 
coordinate for $a=0$ (solid curve), $a=0.1$ (dotted curve), $a=0.5$ (dashed curve), $a=-0.1$ (long-dashed curve) and 
$a=-0.5$ (dot-dashed curve). (d) Variation of shock invariant quantity ($C$) as a function of radial
coordinate  for
inner sonic (solid curve) and outer sonic (dotted curve) point branch when $a=0$; the intersection points 
of two curves indicate
the shock locations which are $x_s=33.77,6.94$. Other parameters are $M=10M_\odot$, $\dot{M}=1$ Eddington rate, $\lambda=3.3$, $x_o=70$ (for
solid, dotted and dashed curves), $x_i=5.6$ (for solid, long-dashed and dot-dashed curves), $\gamma=4/3$.}
\end{figure}
\clearpage

Figure 5a shows the variation of Mach number for a particular set of physical parameter. If the black
hole is chosen to be non-rotating, shock forms in the disk. When the rotation is considered 
keeping other parameters unchanged, the 
solution changes and shock disappears. If the black hole co-rotates, the inner-edge of
disk becomes unstable and the matter does not find any physical path to fall steadily into the black hole. 
On the other hand, if the black hole counter-rotates, there is a single sonic point in the disk
and the matter attains a supersonic speed only close to the black hole and falls into it.
Thus, for the non-rotating and counter-rotating cases, there are smooth solutions of matter passing through
the inner sonic point, while for the co-rotating cases, it is not so.
The physical reasons behind these phenomena can be explained as follows. When the black hole co-rotates,
the angular momentum of system increases, the radial speed of matter may not be able to overcome the centrifugal barrier to fall into
the black hole for that parameter set. If the black hole counter-rotates, the angular momentum of 
system reduces, the centrifugal barrier smears out and matter falls steadily into the black hole. 
It can be mentioned here that, for the other choice of physical parameters (e.g. reducing/increasing the angular 
momentum of accreting matter for co/counter-rotation, changing the sonic locations etc.) the shock may appear again, even for the rotating
black holes (a few such examples are depicted in Fig. 6).
Figure 5b shows the corresponding density profiles in unit of $\frac{c^6}{G^3M^2}$ for a particular accretion rate. 
As the matter slows down abruptly, the density of disk fluid jumps up at the shock location
for a non-rotating case. For the counter-rotating black holes,
the angular momentum of the system is less, thus, to overcome the centrifugal barrier, matter does not need 
to attain a high radial speed away from the black hole. In this situation, the radial 
matter speed is less and the
corresponding density of accreting fluid is high compared to a non-rotating case. We also show the
density profiles for co-rotating, unstable cases. In Fig. 5c, we compare the corresponding temperature
profiles in unit of $10^9$ ($T_9$). The variation of temperature is similar to that of density 
away from the black hole
but is opposite when close to the horizon (as the density varies inversely with the temperature 
at a small $x$ but not so at a large $x$, see Eqn. (\ref{ec})). As the higher counter-rotation
Mach number lowers down, the temperature of accretion disk becomes higher. 

The variation of shock invariant quantity, $C$ is shown in Fig. 5d for the non-rotating case. 
The intersections between the solution of the inner and outer sonic point
branches indicate the shock locations. Since the inner shock location is unstable, the only shock 
forms at an outer radius, $x=37.77$.

\clearpage
\begin{figure}
\epsscale{0.8}
\plotone{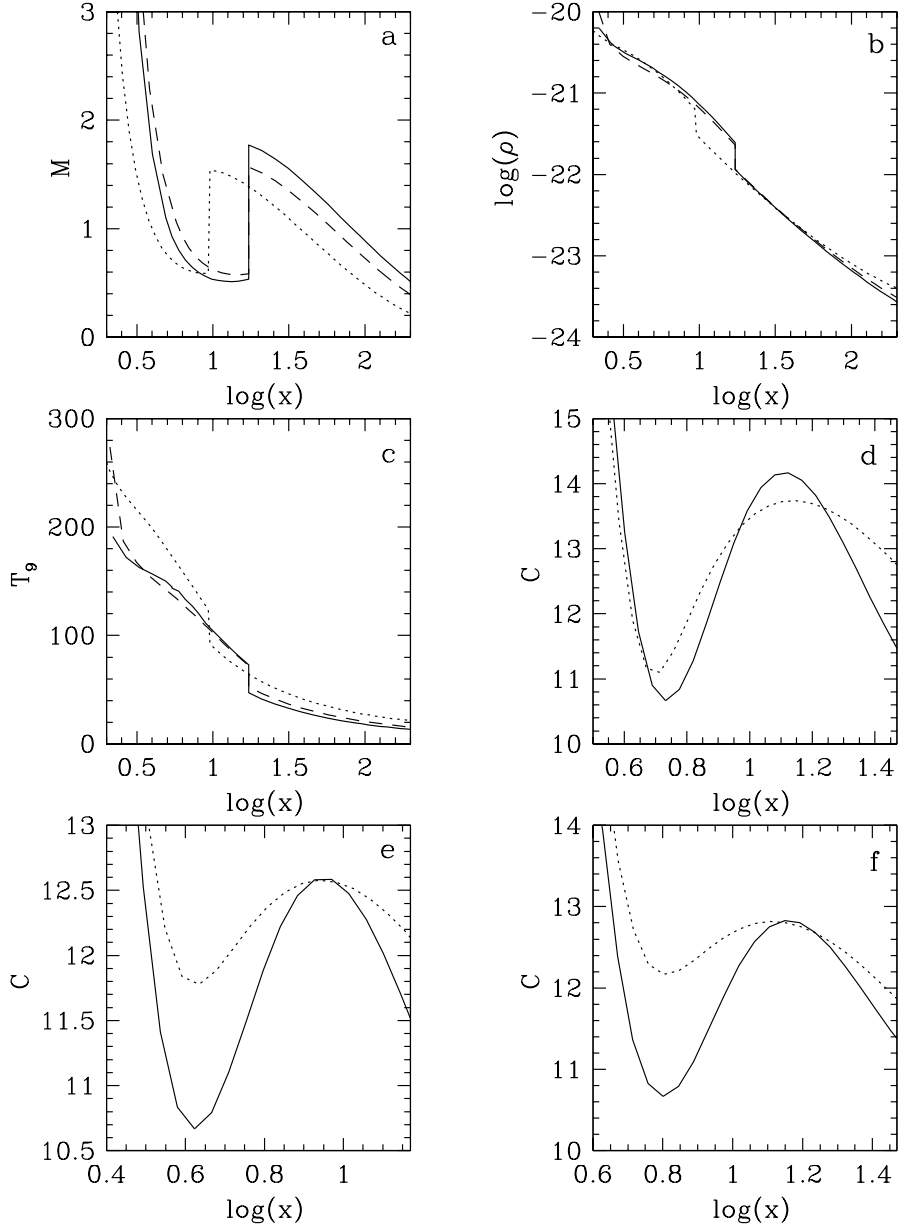}
\caption{
\label{fig6} Variation of (a) Mach no., (b) density and (c) temperature in unit of $10^9$
as a function of radial coordinate when the sets of physical parameters are (i) $a=0.1$, $x_o=94$, $x_i=5.4$,
$\lambda=3.3$ (solid curve); (ii) $a=0.5$, $x_o=43$, $x_i=4.2$, $\lambda=2.8$ (dotted curve) and 
(iii) $a=-0.1$, $x_o=70$, $x_i=6.324$, $\lambda=3.3$ (dashed curve). 
Variation of shock invariant quantity, $C$ as a
function of radial coordinate for the set of (d) parameter (i) when $x_s=17.18,9.2$, (e) parameter (ii) when $x_s=9.51,8.35$ and 
(f) parameter (iii) when $x_s=17.25,13.73$. Solid and dotted curves are for inner and outer sonic branch respectively.
Other parameters are $M=10M_\odot$, $\dot{M}=1$ Eddington rate, $\gamma=4/3$.}
\end{figure}
\clearpage

Figure 6 gives examples where the shock forms in an accretion disk around rotating black hole. 
The variation of Mach numbers for different Kerr parameters are shown in Fig. 6a. 
It reflects that, for a small rotation of black hole ($a=0.1$), 
matter adjusts the sonic locations in such a manner
that the outer ($x_o$) and inner ($x_i$) sonic locations shift in a more outer and more inner region 
respectively with respect to a non-rotating
case, and the shock forms in disk at $x=17.18$. For the counter-rotating case ($a=-0.1$), $x_o$ may remain
unchanged, but $x_i$ has to be shifted outside to form a shock at 
$x=17.25$ (that keeps the shock location (almost) unchanged).
As for the counter rotating cases, marginally stable ($x_s$) and marginally bound ($x_b$) 
orbits shift away from the black hole with respect to a co-rotating one, 
$x_i$ also shifts outside to stabilize the disk.
If the co-rotation of black hole is higher, say $a=0.5$, to form a stable shock, not only the inner sonic 
location has to be smaller, the disk angular momentum ($\lambda$) should also be reduced. 
High rotation of a black hole results in the location of horizon ($x_+$) as well as $x_b$ and $x_s$ 
to shift inwards, and the matter attains a supersonic speed at a point in more inside of the
inner edge of the disk to fall into the black hole.
Thus all the phenomena, like sonic transitions, shock formations, etc., shift inside.
Moreover, for the stable solution, the angular momentum of the system can not be very high, otherwise
matter will be unable to overcome the centrifugal barrier. Thus, for the high rotation of a black hole,
like $a=0.5$, $\lambda$ has to be reduced to form a shock (at $x=9.51$) 
and stabilize the disk. However, for $a=-0.5$ and onwards,
there is no sub-Keplerian flow for which a shock may form. For the higher counter-rotation 
of a black hole, the angular momentum of system reduces. 
Since a higher centrifugal barrier is essential to form a shock, at the higher counter-rotation,
shock completely disappears in the accretion disk. 
Figures 6b and 6c show the corresponding variations of density and
temperature respectively. In Figs. 6d,e,f, we show the corresponding variation of shock invariant quantity $C$ for the inner and
outer sonic point branch, when $a=0.1,0.5,-0.1$ respectively. Again it becomes clear that, for the faster
rotation of a black hole, the shock location shifts inside the disk.

\clearpage
\begin{figure}
\epsscale{0.9}
\plotone{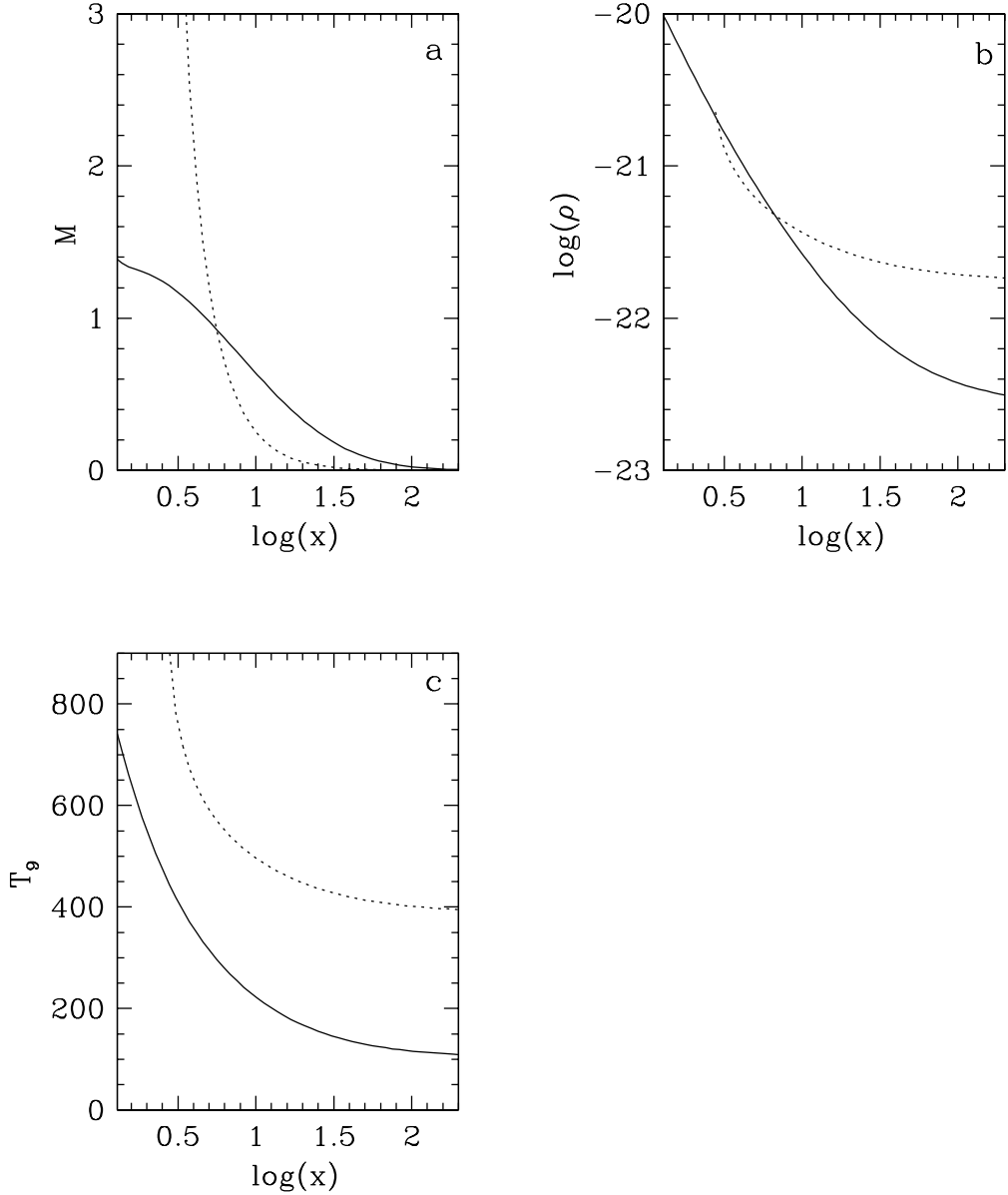}
\caption{
\label{fig7} Variation of (a) Mach no., (b) density and (c) temperature in unit of $10^9$ as a function of radial
coordinate for the sets of parameters (i) $a=0.998$, $x_i=5.6$, $\lambda=1.8$ (solid curve) and (ii) $a=-0.998$,
$x_i=5.6$, $\lambda=3.3$ (dotted curve). Other parameters are $M=10M_\odot$, $\dot{M}=1$ Eddington rate,
$\gamma=4/3$.}
\end{figure}
\clearpage

In Fig. 7, we show the fluid properties for $a=\pm 0.998$. At a very high rotation of black hole,
the possibility to form a shock in accretion disk totally disappears. For the co-rotating case,
the disk cannot be stable unless the angular momentum of accreting matter is very low 
(as we have already mentioned that the high angular momentum of the system makes the disk unstable). 
Therefore, there is no scope to form a shock, as to its formation, matter
needs a significant centrifugal barrier as well as a high angular momentum of the accreting fluid.
Here we choose, $\lambda=1.8$ and the matter falls almost freely. For 
the counter-rotating case, the matter angular momentum need not be low to stabilize the disk. 
However, in the 
sub-Keplerian accretion disk, $\lambda$ cannot attain a high value to increase the
angular momentum as well as the centrifugal barrier of a system to form a shock. Thus, shock also disappears
for a high counter-rotating case. Figure 7a shows that $x_b$ and $x_s$ shift outwards for the retrograde orbit 
and the matter attains a supersonic speed at an outer radius than the case of a direct orbit of the disk. 
However, far away from black hole, the radial velocity is less (as it needs to 
overcome a little centrifugal barrier) than that of co-rotating case. 
Figure 7b shows the variation of corresponding density. Less velocity implies a high rate of
piling up of matter at a particular radius that produces a high density in the disk and thus, the density 
behaves in an opposite manner to Mach number. Figure 7c shows the variation of virial
temperature in accretion disk. The temperature is low for the co-rotation and high for the counter-rotation. 
The physical reason behind it is as follows. For a co-rotation, the net speed as well as the kinetic 
energy of matter is less, producing 
 less temperature. On the other hand, for a counter-rotating case, the situation is opposite.
Although we show the virial temperature of the disk, in reality several cooling processes may take place 
that can reduce the disk temperature upto a factor of two orders.

\section{Discussion and Conclusion}

Here, we have studied the stability of accretion disk around rotating black holes in a 
pseudo-Newtonian approach. Following Paper-I, we have incorporated the general relativistic 
effects (according to Kerr geometry) to the accretion disk in such a manner that all the essential relativistic 
properties, like, the locations of the marginally bound and stable orbits, black hole horizon and 
mechanical energy of the accreting matter are reproduced in the disk for various values of Kerr parameters.
Thus we can say, our method is a 'pseudo-general-relativistic'. 
As our interest is to check how the rotation of a black hole solely affects the disk structure,
we have chosen the inviscid fluid in accretion disk. Most of the earlier studies on structure and
stability analysis of the accretion disk have been done around a non-rotating black hole 
(Schwarzschild geometry). As no cosmic
object is static, it is very important to consider the effect of rotation of the black hole, particularly
for the discussion of an inner edge of accretion disk. We have found that when the rotation of a black hole
is incorporated, the valid, known disk parameter region of a non-rotating black hole is 
dramatically affected. As a result, the location of various sonic points and shock (if any) get shifted severely
and influence the disk structure. Therefore, we can conclude that in the study of an accretion disk,
the rotation should be considered. Any related conclusion depends on the rotation parameter of a
black hole. So, we can say that the known physical parameter regime (Chakrabarti 1989, 1990, 1996a)
to form a shock, sonic locations etc. in the disk around a non-rotating black hole is not global. The mentioned
disk properties and phenomena not only get modified and/or shifted in location at the disk, but also sometimes 
completely disappear.
In our study, we have chosen the Kerr parameter from $a=0$ to the case of an extremely high 
rotation as $a=0.998$ and thus, we now have a clear global picture of the accretion disk. 

Though, in the literature, there are a few works on the accretion disk
around Kerr black holes (e.g. Chakrabarti 1996b,c; Gammie \& Popham 1998; Popham \& Gammie 1998), 
those do not concentrate on the
comparative study of a rotating and non-rotating black hole and the instability of disk induced 
by the rotation of black hole. Chakrabarti (1996b,c) mainly concentrated on fast rotating black holes and showed
solution topology in full Kerr geometry. He does not show, how the solution varies with the change
of rotation of the black hole, keeping unchanged other physical parameters. Thus, it is not clear
from those works, how the Kerr parameter affects the disk solution solely. On the other hand,
Gammie \& Popham (1998) and  Popham \& Gammie (1998) do not discuss the effect of rotation 
on the sonic points globally, which is shown here in Figs. 1-4. However, the fluid dynamical results
achieved by them tally with the present work. For example, they 
(Gammie \& Popham 1998;  Popham \& Gammie 1998) showed, 
the effect of non-zero Kerr 
parameter is dominant close to the black hole, that is also reflected from
Fig. 4 of the present work, which shows the outer sonic point branches are unchanged, while the inner ones 
are shifted.
They also showed, the (inner) sonic points move to smaller (larger) radii as co-rotation (counter-rotation)
increases in magnitude. In our work, similar features come out from Figs. 1,2,3 and 6.
Thus, we can conclude once again that, the pseudo-potential prescribed in Paper-I can reproduce the 
general relativistic properties in accretion disk. Unlike the works by
Gammie \& Popham (1998) and  Popham \& Gammie (1998), here we have chosen inviscid flow to understand
the sole effect of rotation of the black hole on the fluid and corresponding disk 
structure. Finally we can conclude that the effect is significant.

Here, we have carried out the stability analysis of accretion disk by choosing the black hole 
to be rotating.  Prasanna \& Mukhopadhyay (2002) have worked on the stability of accretion disk around rotating
compact objects (but not for black holes) in a perturbative approach. They incorporated the rotational effect of 
central object in a disk indirectly, by the inclusion of Coriolis acceleration term. Here, the compact object is
black hole and its rotational effect is brought from the Kerr metric only.
As we have a very self-consistent pseudo-potential (Paper-I), it has been now easy to study the accretion phenomena
around a rotating black hole. 

From the global analysis of sonic points, we have established that the disk becomes unstable 
for a higher co-rotation at a particular angular momentum of the accreting matter. On the other hand,
to form a shock, accretion disk must have a stable inner sonic point. Thus
for higher $a$ (Kerr parameter), shock is unstable,
as the inner region of disk itself is not stable. In that regime, shock may disappear by any disturbance
created on the accreting matter and thus the matter may not get a steady inner sonic point to
form a stable inner disk. For the counter-rotating cases, the angular momentum of 
the system reduces and the matter falls to the black hole more rapidly. As there is 
no significant centrifugal barrier to slow
down the matter, the possibility of shock reduces again. For a very high counter-rotation,
shock disappears completely. Thus we can conclude that 
the parameter region where the shock is expected to form (Chakrabarti 1989, 1990, 1996a) for non-rotating
black holes, gets affected for the rotating ones. For a positive $a$, shock might have been formed 
for a lesser value of the angular momentum of accreting matter. On the other hand, for a negative $a$,
to form a shock, the angular momentum of accreting matter should have a larger value than that
of non-rotating cases. As we are studying the sub-Keplerian accretion flow, the angular
momentum of matter cannot be unlimitedly high 
(e.g., for a non-rotating black hole, $\lambda\lsim 3.6742$ at last stable orbit). 
Thus, to form a shock, matter cannot attain the angular momentum beyond a certain limit. 
Also for a highly co-rotating black hole, to stabilize the disk, $\lambda$ has to be reduced by some physical process,
otherwise the radial speed could not overcome the centrifugal pressure to form a steady inner edge of the disk.
Therefore, the disk would not be transonic and could be unstable in this situation.
Because of a reduced $\lambda$, centrifugal force as well as shock strength reduces, 
as a result, the possibility of a shock is also reduced for both the co-rotating and counter-rotating 
black holes. However, at an intermediate co-rotation, 
the parameter region of accretion disk enlarges having three sonic points
and there is a possibility of a stable shock formation. On the other hand, for a counter-rotation, the valid
parameter region with three sonic points at the disk shortens and thus the possibility of shock reduces.

Now the question may arise, how would a flow behave in the region where sonic points disappear
because of the higher rotation of black hole? It can be thought as, matter would try to pass through
same sonic point as of non-rotating (or slowly rotating) case. Because of
the absence of sonic point, the searching procedure for the true sonic location should give rise to
an unstable solution. Only the time-dependent simulation of accretion disk can tell whether this idea is 
correct or not. Overall we can say that the rotation of a black hole affects the sonic locations 
of an accretion disk
which are directly related to the formation of stable disk structure. 
Therefore, the stability of an accretion disk
is strongly related to the rotation of central black hole. 

It has also been confirmed that the potential given in Paper-I is very well applicable 
for the global solution and corresponding fluid dynamical study of the accretion disk around a 
rotating black hole. The set of a full
general relativistic disk equations is very complicated for a Kerr black hole, particularly, 
when the accreting fluid is 
chosen to be viscous. Thus, in the study of a relativistic accretion disk, this pseudo-potential 
is very effective to avoid the complexities of full general relativistic fluid equations. 
The next step should be the study of a viscous accretion disk using this general pseudo-potential. 

\begin{acknowledgements}

I am grateful to my friend Subharthi Ray (of IF-UFF) for careful reading the manuscript
and amending the grammatical errors.

\end{acknowledgements}

{}

\end{document}